\documentclass[useAMS,usenatbib]{mn2e}
\usepackage{Times}

\usepackage[T1]{fontenc} 
\usepackage{ae,aecompl}
\usepackage{amsmath}
\usepackage{amsfonts}
\usepackage{amssymb}
\usepackage{graphicx}
\usepackage{natbib}

\title[On supercycle lengths of active SU~UMa stars]{On supercycle lengths of active SU~UMa stars}
\author[M. Otulakowska-Hypka and A. Olech]
{M.~Otulakowska-Hypka$^{1}$\thanks{E-mail:magdaot@camk.edu.pl}, A.~Olech$^{1}$\\
$^{1}$N. Copernicus Astronomical Center, Polish Academy of Sciences, ul. Bartycka 18, 00-716 Warsaw, Poland}

\begin{document}

\date{Accepted yyyy Month dd. Received yyyy Month dd; in original form yyyy Month dd}

\pagerange{\pageref{firstpage}--\pageref{lastpage}} \pubyear{2002}

\maketitle

\begin{abstract}
We performed a detailed analysis of extensive photometric observations of a sample of most active dwarf novae, that is SU~UMa stars which are characterised by supercycle lengths shorter than 120~days. 
We found the observational evidence that supercycle lengths for these objects have been constantly increasing over the past decades, which indicates that their mean mass transfer rates have been decreasing during that time. This seems to be a common feature for this type of stars. 
We present numerical results in each case and estimate time scales of future development of these systems.
This study is important in the context of evolution of dwarf novae stars and perhaps other cataclysmic variables. 
\end{abstract}


\section{Introduction}

As it was pointed out in one of the papers of \citet{1997Paczynski}, massive professional photometric surveys are today sources of many, often unexpected, discoveries (e.g. the ASAS project of \citet{1997Pojmanski} or the Kepler mission of \citet{2010kepler}).
On the other hand, databases collecting amateur photometric measurements are growing rapidly, as telescopes and CCD detectors are easily accessible nowadays \citep{2011Patterson, 2012Szkody, 2012Waagen}.
This gives us access to a number of valuable open astronomical archives, both professional and amateur, with a richness of data gathered during the past decades. We have now a great opportunity to trace photometric behaviour of a lot of variable objects, for instance dwarf novae stars. 

Dwarf novae are cataclysmic binaries with a white dwarf as a primary and a low mass main sequence star, filling its Roche lobe, as a secondary. Due to the transfer of mass from the secondary onto the non-magnetic primary, an accretion disk is formed. 
Such a mass transfer leads to outbursts in dwarf novae, as they are caused by sudden gravitational energy release due to accretion of this material onto the white dwarf. Dwarf novae with shortest orbital periods ($P_{orb} < 2.5 ~\rm{h}$) are called SU~UMa stars. They are characterised by the presence of superoutbursts, next to normal outbursts, which are about $1~\rm{mag}$ brighter and about ten times less frequent than normal outbursts \citep{2001Hellier, 2003Warner}.

The length of the supercycle, i.e. the time between two successive superoutbursts, is one of the most fundamental properties of SU~UMa stars and is specific for each of them. The nature of this feature is still uncertain. 
Nowadays there is an open debate about the cause of superoutbursts.
The thermal-tidal instability (TTI) model \citep{1996Osaki} suggests that the sypercycle length ($P_{sc}$) is set by the mass transfer rate ($\dot{M}_{tr}$). According to this scenario, the observed $P_{sc}$ is inversely proportional to $\dot{M}_{tr}$. The TTI model shows that normal outbursts remove less material from the disk than it has been accreted since the last normal outbursts. This makes the disk grow during a number of normal outbursts. When the disk reaches the critical radius ($r \sim 0.46 a$, with $a$ being the binary separation), it becomes eccentric and tidal effects occur, which causes superoutburst, during which the disk is shrinking again.
Although the TTI model is commonly accepted, it still seems to need some improvements. 
For instance the TTI simulations which reproduced the light curve of one of the most active dwarf novae, ER~UMa star, required an artificial increase of the mass transfer rate by the factor of ten in comparison to values expected from the theory based on gravitational radiation \citep{1995Osaki}.
Another example is the fact shown by \citet[and references therein]{2012Cannizzo} that embedded precursors of superoutbursts are present also in systems with long orbital periods. This argues for a more general model, which is not restricted to the mass ratio of $q<0.25$, as the TTI model.
Recently \citet{2012Osaki} presented evidence in favor of this model.
Although their analysis was based on variations of the negative superhump period of a single dwarf nova, V1504 Cyg, they claimed that the cause of superoutbursts is finally revealed.
However, \citet{2013Smak} showed that this object cannot be considered as a representative for all systems which show negative superhumps. He also presented a number of arguments which suggest that superoubursts are caused by strongly enhanced mass transfer (EMT) rate \citep{2008Smak}. 

Recent observations \citep{2012MOH} surprisingly showed that the supercycle length of one of the most active dwarf novae, IX~Draconis, has been increasing with a constant rate since the last twenty years. This is interesting in the context of evolution of such systems. Encouraged by this fact, we decided to investigate the same issue for other well-observed active objects of the SU~UMa type. 

The paper is arranged as follows. In Section~\ref{sec-data} we give information on the data used in this research. The analysis is presented in Section~\ref{sec-analysis}. Results together with discussion are given in Section~\ref{sec-res}. We summarize the main conclusions in Section~\ref{sec-conc}.

\section{Data}
\label{sec-data}

\begin{table*}
\caption{List of objects with information on the used databases in this study.}
\begin{tabular}{c|c|c|c|c|c|c|c}
\hline
Object		& 	ASAS	&	MEDUZA	&	AAVSO	&	AFOEV	&	BAAVSS	&	Kepler	& extra data \\
\hline
DI~UMa 			&		&	x		&	x		&	x		&	x		&		&	\citet{2009Rutkowski} \\
ER~UMa 			&		&	x		&	x		&	x		&	x		&		&	\citet{1995Kato}	\\
RZ~LMi 			&		&	x		&	x		&	x		&	x		&		&	\citet{2008Olech}	\\
SS~UMi 			&		&	x		&	x		&	x		&	x		&		&	\citet{2006Olech}	\\
V1159~Ori 		&	x	&	x		&	x		&	x		&	x		&		&		\\
V1504 Cyg		&		&	x 		&	x 		&	x 		&	x 		&	x	&		\\
V344 Lyr		&		&	 		&	x 		&	x 		&	x 		&	x	&		\\
V503~Cyg 		&		&	x		&	x		&	x		&	x		&		&		\\
YZ~Cnc 			&	x	&	x		&	x		&	x		&	x		&		&		\\
\hline
\end{tabular}
\label{tab-objects}
\end{table*}

We searched accessible archives of amateur astronomical observers and automatic professional surveys to create joint and as complete light curves as possible. In a few cases we also have an additional data from our previous observational campaigns and from other observers.
The relevant photometric data was obtained from the following databases:
the ASAS\footnote{http://www.astrouw.edu.pl/asas/} project \citep{1997Pojmanski}, 
the MEDUZA\footnote{http://var2.astro.cz/} project of the Variable Star and Exoplanet Section of the Czech Astronomical Society \citep{2010meduza},
the Kepler mission\footnote{http://archive.stsci.edu/kepler/} \citep{2010kepler},
and the databases with amateur observations: 
AAVSO\footnote{http://www.aavso.org/}, 
AFOEV\footnote{http://cdsarc.u-strasbg.fr/afoev/},
and BAAVSS\footnote{http://www.britastro.org/vss/}.

At first, we intended to perform the analysis for most of the well-observed (which often means the brightest) active SU~UMa stars. However, we found not enough data for the following objects: BF~Ara, BK~Lyn, BR~Lup, CI~UMa, IX~Dra, MN~Dra, NY~Ser, SDSS~J210014, and UV~Gem, to reach any unambiguous conclusions. 
We selected only those targets which have really \textit{good} time coverage of their light curves, i.e. those, for which we were able to easily distinguish superoutbursts from the rest of the signal. The chosen objects have light curves spreading over at least twenty years. 
In Table~\ref{tab-objects} we present the final list of objects together with information on the used databases for each star.

\begin{figure}
\begin{center}
\includegraphics[width=0.38\textwidth, angle=270]{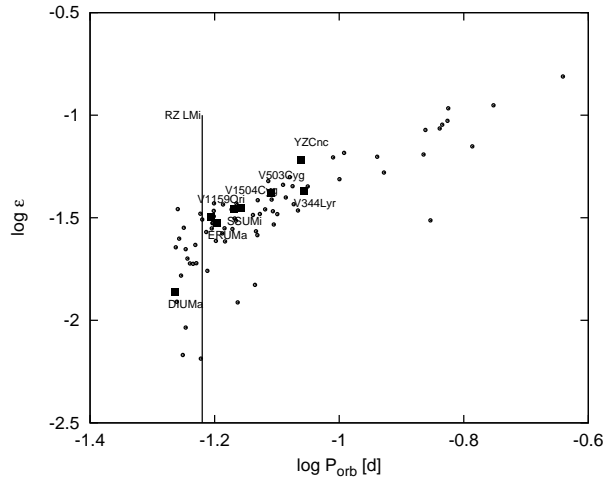}
\caption{Relation between the orbital period-superhump period excess ($\epsilon$) and the orbital period ($P_{orb}$) for dwarf novae stars. The objects used in this research are shown by squares. The line indicates possible positions of RZ~LMi star.}
\label{fig-evol}
\end{center}
\end{figure}

Although our selected objects are all of the SU~UMa type and they are located below the period gap in the orbital period distribution for dwarf novae, they are diverse in terms of the evolutionary status. 
Figure~\ref{fig-evol} shows the relation between the orbital period-superhump period excess ($\epsilon$) and the orbital period ($P_{orb}$) for dwarf novae stars, noticed first by \citet{1984Stolz}, where we marked by squares positions of the selected objects. 
Thanks to the dependence between $\epsilon$ and the mass ratio ($q$) for dwarf novae stars
reported by \citet{1998Patterson}:
\begin{equation}
\epsilon = \frac{0.23 q}{1+0.27q},
\end{equation}
we can use the relation between $P_{orb}$ and $\epsilon$ as an excellent plane to examine evolution of these stars, since the mass ratio decreases with time due to the mass loss from the secondary.
In Fig.~\ref{fig-evol} small points are known dwarf novae objects, from \citet{2011Olech}. 
It was impossible to mark the exact position of one of the stars, RZ~LMi, since its orbital period is unknown. 
However, based on the value of its superhump period, we know that it must be located somewhere on the solid line, which is shown in this figure.

\section{Analysis}
\label{sec-analysis}

To create final joint light curves from all data sources listed in Tab.~\ref{tab-objects}, we decided to use only the certain data points, without upper limits or visual observations. 
For all the data sources, with the exception of the Kepler data, the merger of light curves was simple and straightforward. The only thing to do was to bring all the measurements into the common final time system, the HJD.
For light curves from the Kepler mission this process was different. Here we had to transform the flux
into magnitudes. 
For this we used the Pogson's equation plus an arbitrary chosen constant shift, since we had no reference point. 
We did this step very carefully, but it is still possible that there exists some error between magnitudes from the Kepler part of the data and the rest of data sources.
Even if this was true, it would be insubstantial, since our study concerns time series analysis, not the brightness.
In the case of the Kepler data the time was originally given as BJD-2454833.0. 
The rest of the data was in the HJD. We did not have to convert one time system to another because the difference between these two standards is of the order of a few seconds \citep{2010Eastman}, which is a few orders of magnitude below our precision of determination $P_{sc}$. 

As a result we obtained a huge data sets with a number of photometric measurements of the order of tens of thousands for each of the objects.

\begin{figure*}
\includegraphics[angle=270,width=0.7\textwidth]{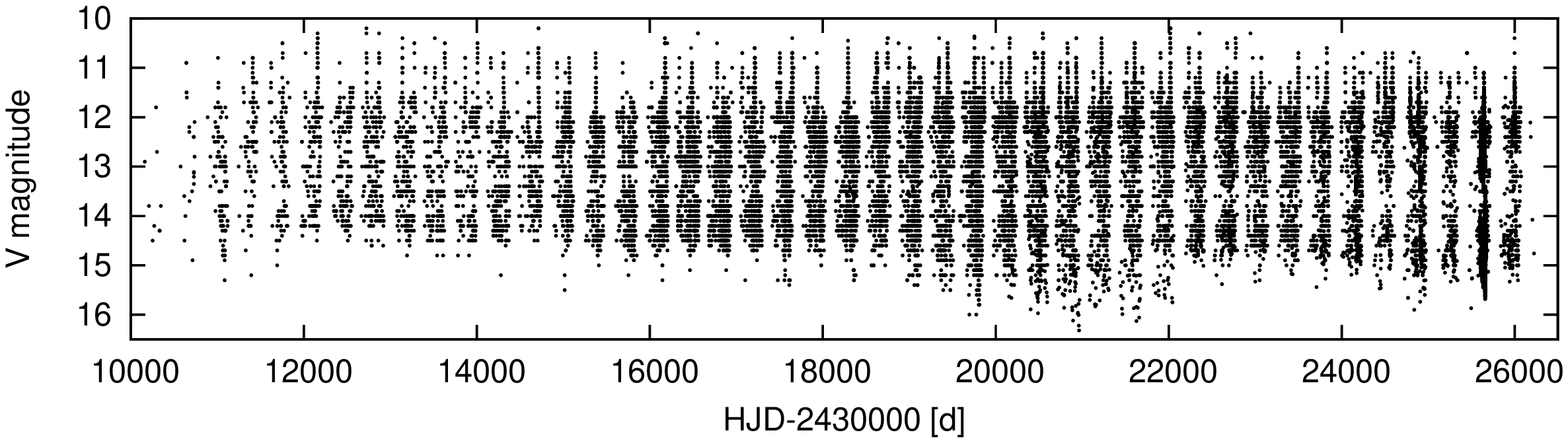}\\
\includegraphics[angle=270,width=0.7\textwidth]{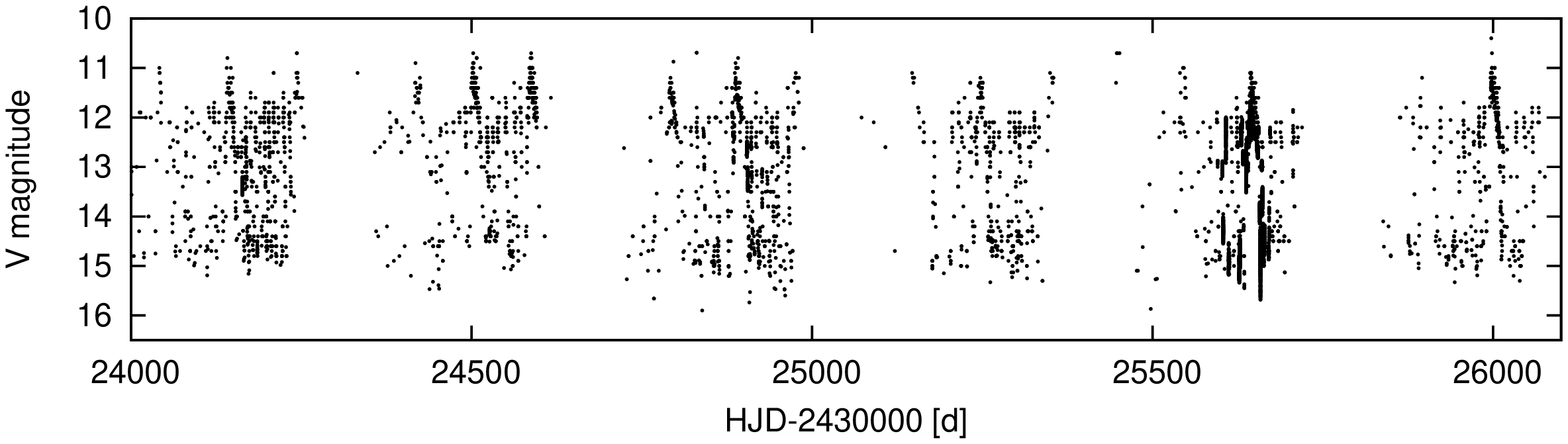}
\caption{Example of a final light curve of YZ~Cnc used in the further analysis (top). 
Zoom in the most recent part of this light curve where superoutbursts are clearly visible (bottom).}
\label{fig-lc}
\end{figure*}

In Fig.~\ref{fig-lc} we show as an example the final light curve for YZ~Cnc star, which was used in the further analysis. The obtained light curves of all the objects are not uniform. 
There is a yearly trend present in all of them, with the exception of the Kepler data, caused by the variable observational conditions. 
In addition, the density of observations increases with time, as the number of amateur observers is constantly growing.

\begin{figure}
\includegraphics[angle=270,width=0.48\textwidth]{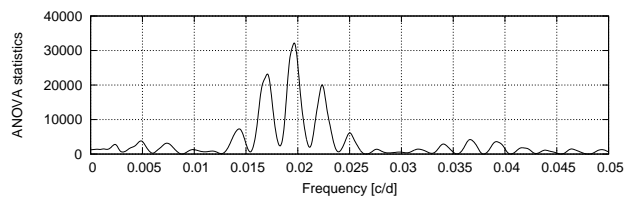}
\caption{An example of a resulting periodogram. This is the case for the last time bin of ER~UMa star. The most prominent peak here corresponds to the $P_{sc}=50.9\pm0.5$~d.}
\label{fig-per}
\end{figure}

For this unevenly sampled data, we decided to use the ANOVA software \citep{1996Schwarzenberg} for the time series analysis. 
First, we divided each of these light curves into five bins of equal time range. For each of the bins we searched for the most prominent peak in the corresponding power spectrum obtained from ANOVA. 
The data suffered from aliasing because of the under-sampling. 
However, we were able to derive values of $P_{sc}$ for vast majority of the time bins for all of the objects. 
These values correspond to the most dominant peaks in a reasonable range of frequencies of periodograms. 
An example of such power spectrum is presented in Fig.~\ref{fig-per}.

The obtained value of supercycle length for a given time bin is naturally not exactly equal to each time period between each pair of successive superoutbursts during the time covered by the time bin, because not all of them are strictly regular. There are some subtle differences between the $P_{sc}$ measured on short time scales, as was already shown many times in the literature, for instance by \citet{2005Antonyuk} and \citet{2012Zemko}. However, our analysis aimed at examination of the overall behaviour of the $P_{sc}$, i.e. on long time scales of the order of decades. Thus, based on the ANOVA statistics, we derived one value of $P_{sc}$ for each time bin, which corresponds to the range of observation time from a few years to one decade, depending on the object. All the results are presented in the next section, and the exact ranges of time bins for all stars are given in Table~\ref{tab-results}.

\section{Results and discussion}
\label{sec-res}

\begin{figure*}
\includegraphics[angle=270,width=0.3\textwidth]{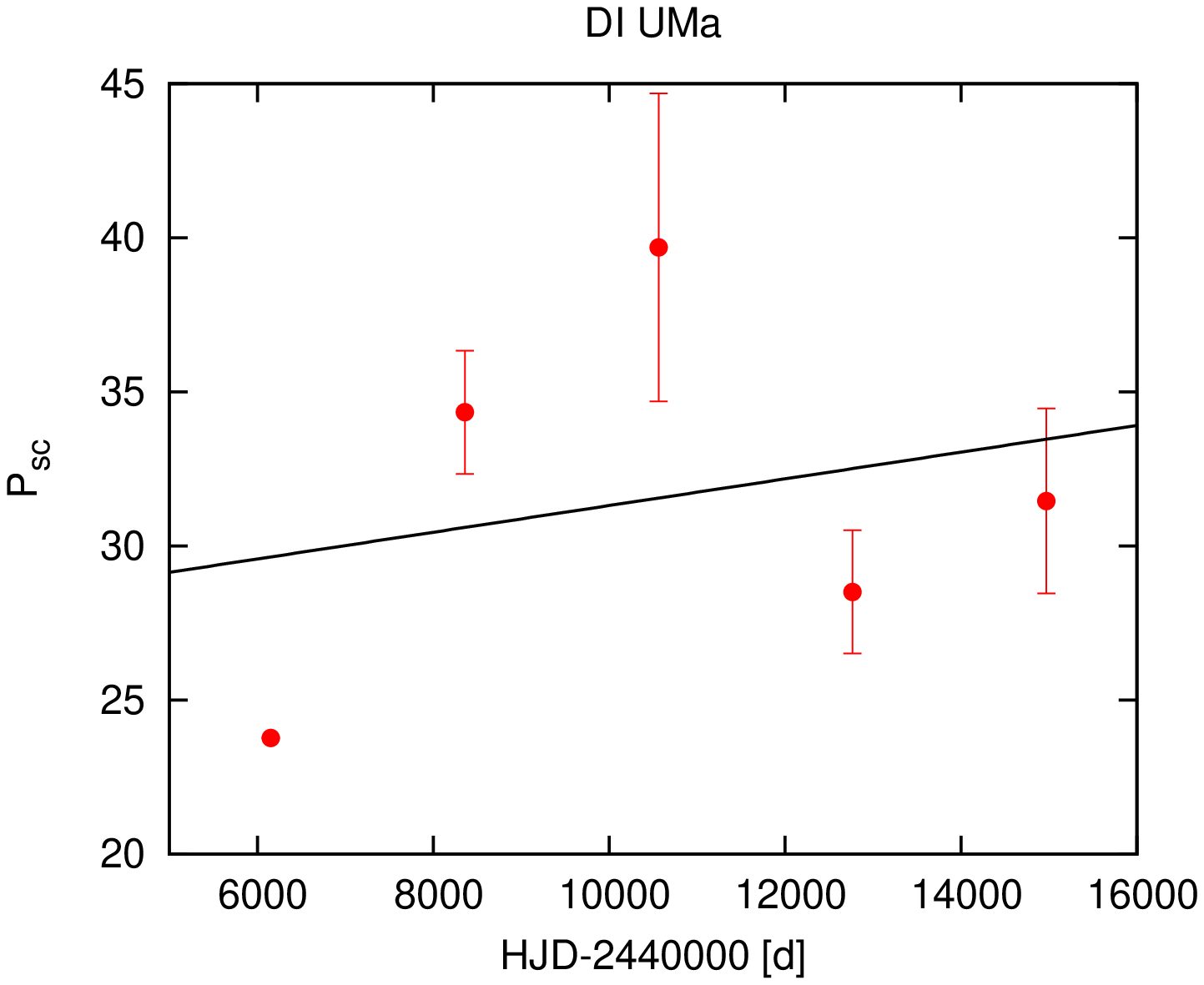}
\includegraphics[angle=270,width=0.3\textwidth]{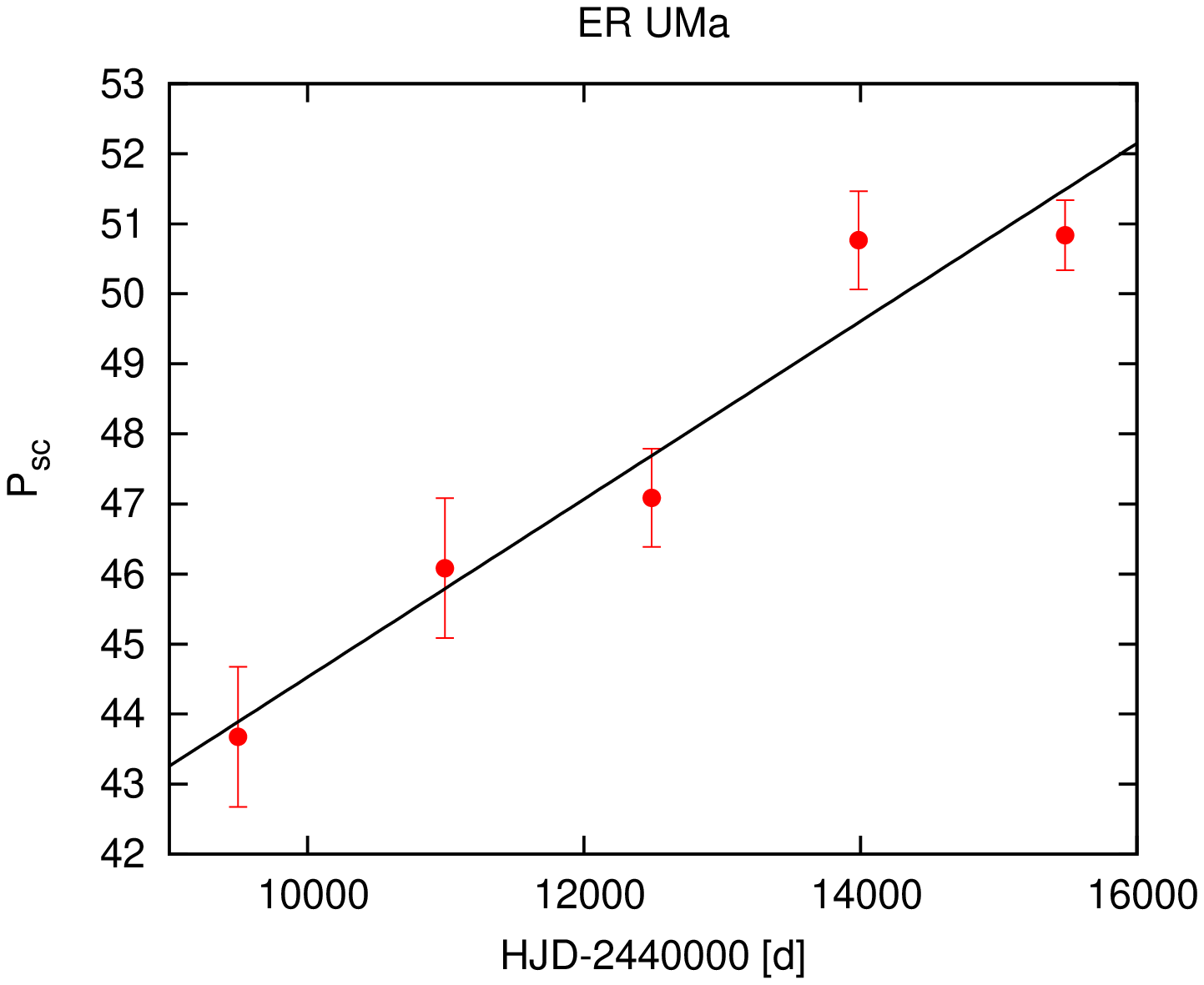}
\includegraphics[angle=270,width=0.3\textwidth]{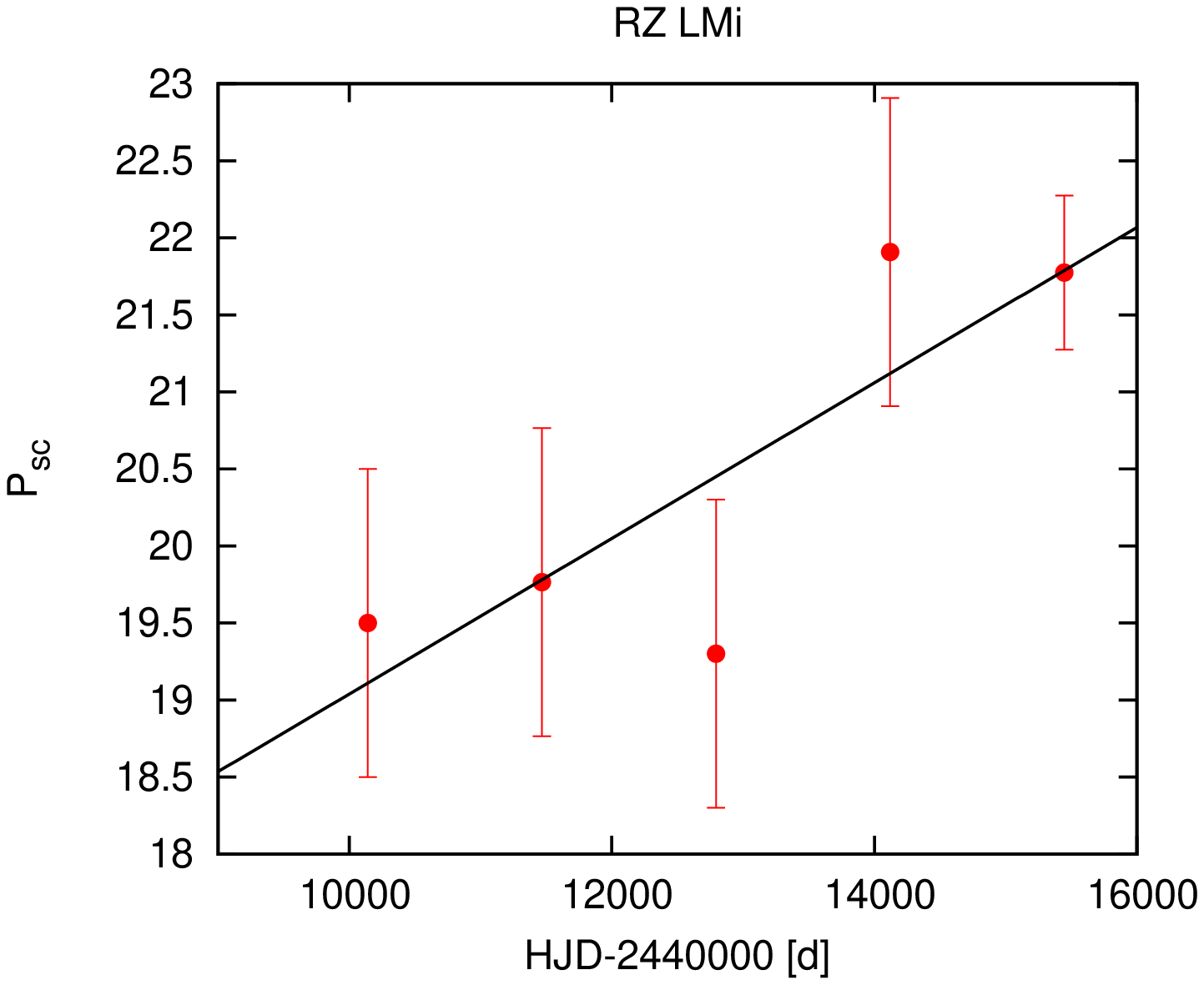}
\includegraphics[angle=270,width=0.3\textwidth]{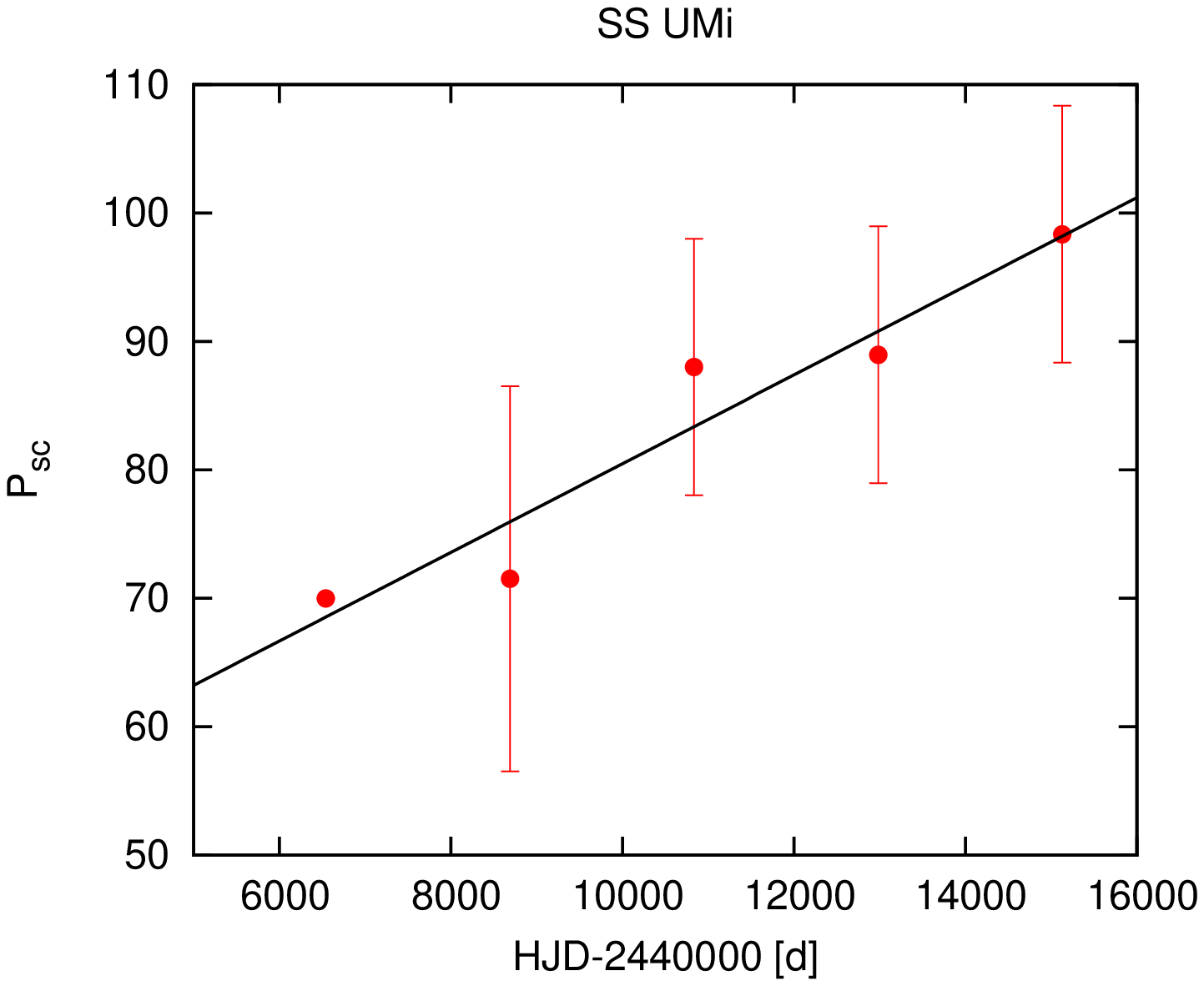}
\includegraphics[angle=270,width=0.3\textwidth]{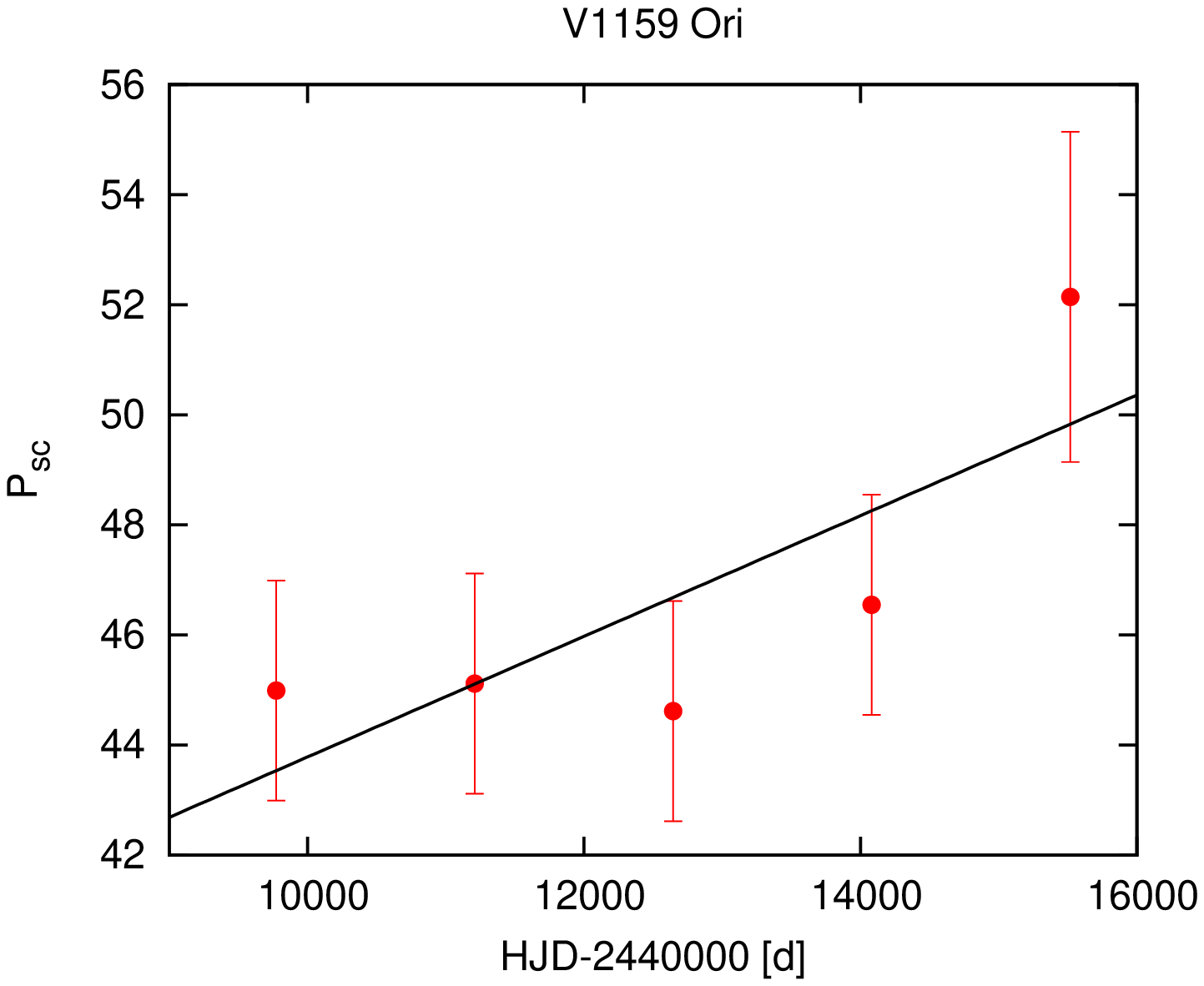}
\includegraphics[angle=270,width=0.3\textwidth]{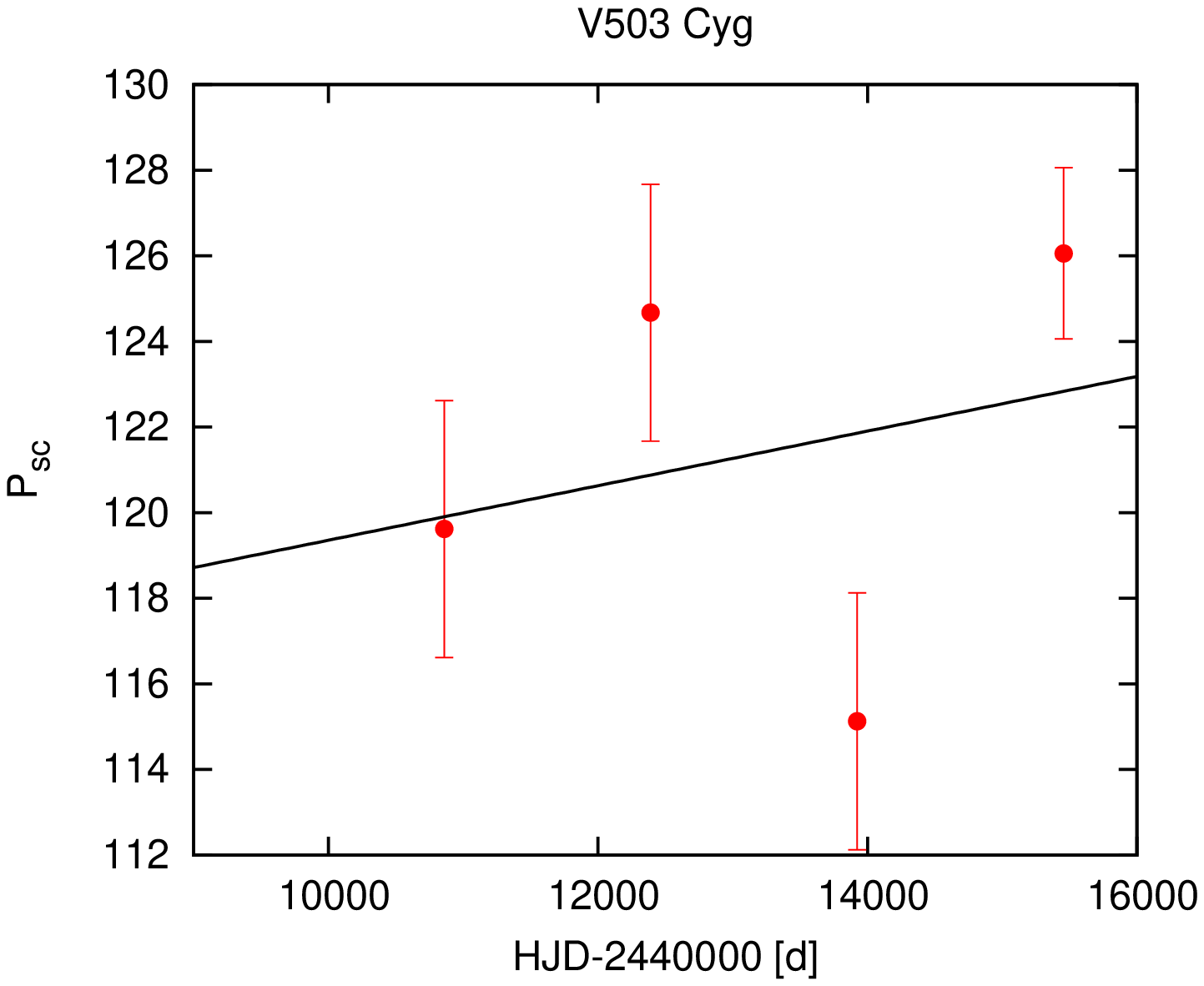}
\includegraphics[angle=270,width=0.3\textwidth]{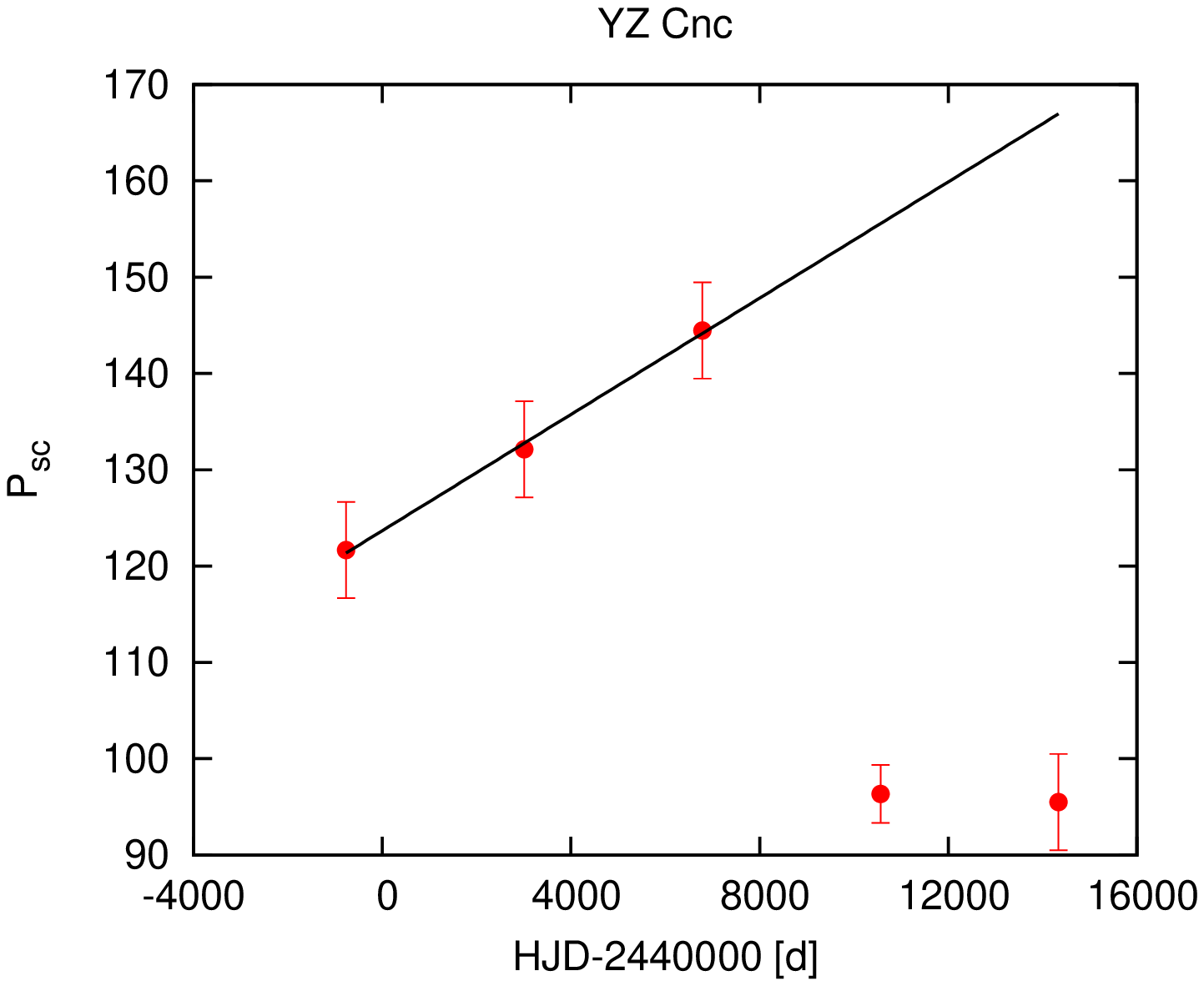}
\caption{Results from ANOVA which show \textit{increasing} supercycle lengths. Data points without uncertainties are possible but not certain due to poor light curves coverage.}
\label{fig-res-anova}
\end{figure*}

\begin{figure*}
\includegraphics[angle=270,width=0.3\textwidth]{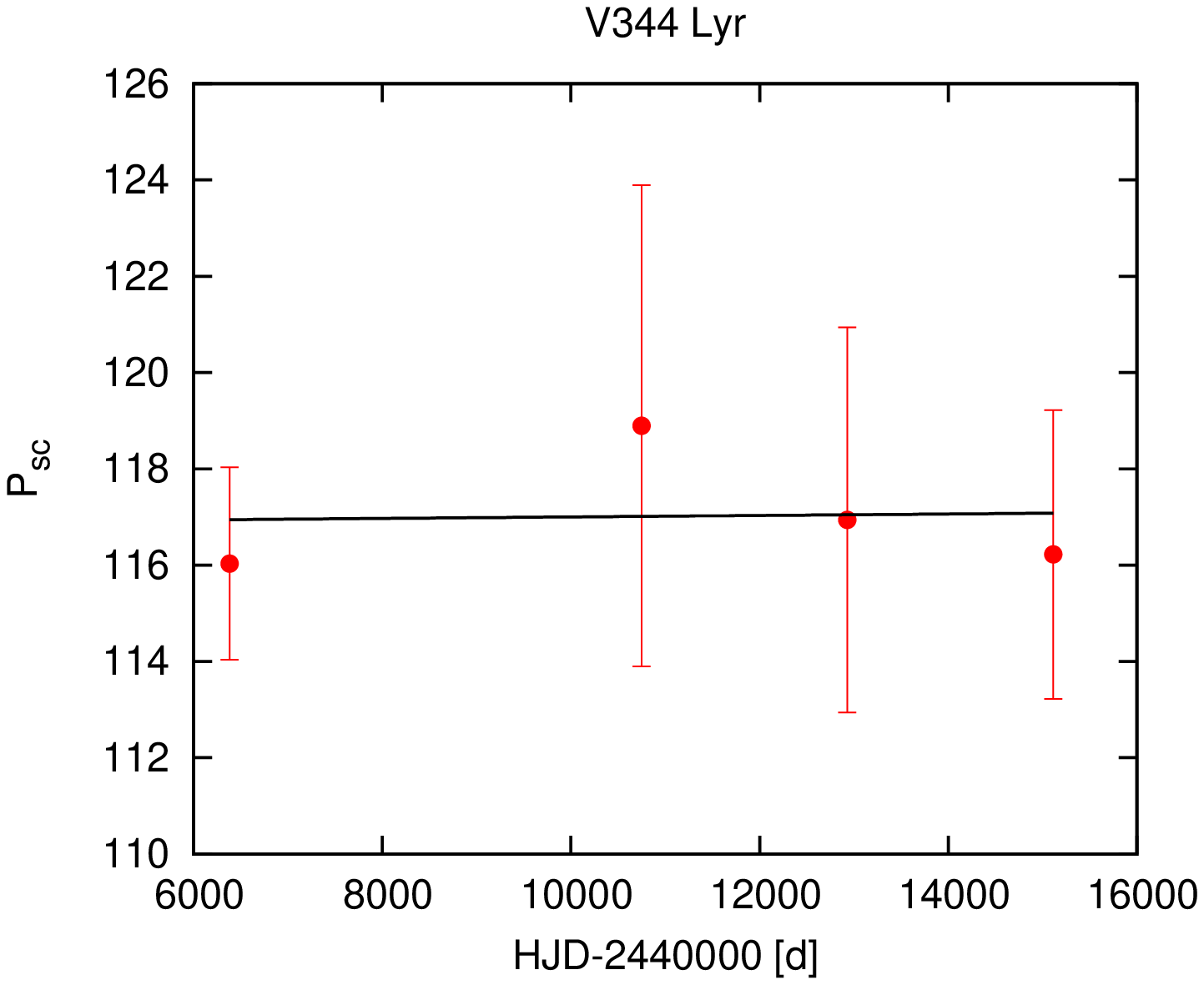}
\includegraphics[angle=270,width=0.3\textwidth]{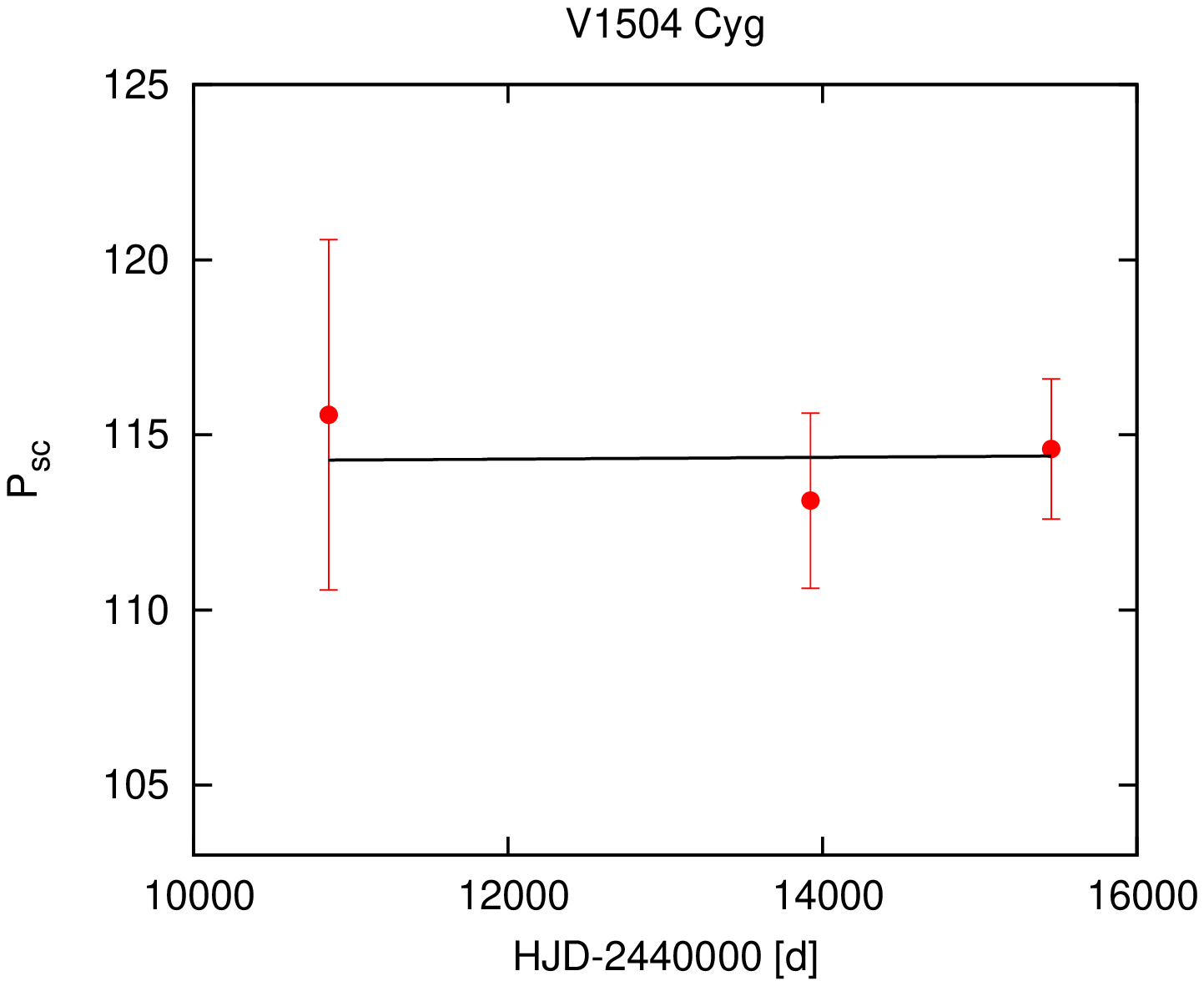}
\caption{Results from ANOVA which show \textit{constant} supercycle lengths. Due to the light curves quality, for some of the time ranges the analysis was impossible. We show only the available measurements.}
\label{fig-res-constant}
\end{figure*}

In Fig.~\ref{fig-res-anova} and \ref{fig-res-constant} we show the obtained evolution of supercycle lengths for all of the analysed objects.
Unfortunately, the quality of the rest of the data did not allow us to investigate more objects, so the statistics is poor. 
Seven out of nine examined targets show clearly growing supercycle lengths during the past decades (Fig.~\ref{fig-res-anova}). 
The only two examples which seems to have a constant value of $P_{sc}$ are the ones with the poorest time coverage of the observations, before the Kepler data, which could have influenced the results (Fig.~\ref{fig-res-constant}). 

The behaviour of the $P_{sc}$ of YZ~Cnc (the last one in Fig.~\ref{fig-res-anova}) is a puzzle.
First of all, it has the highest value of its period among all of the objects. 
Secondly, there is a peculiar drop in the $P_{sc}$ between the third and the fourth time bin, which looks like a transition from the increasing to constant supercycle length.
It is possible that this is only a short time scale fluctuation, like for instance in the case of V503~Cyg.
However, this drop is significant in comparison to the uncertainties of a single measurement, unlike in the case of all the rest of the objects, which have fluctuations of the order which is about one magnitude smaller.
Since this effect is so strange and unique, we checked it twice, and we claim that this is certainly not an artificial result. 
Because of the fact that the value of the supercycle length for YZ~Cnc is the highest one, we cannot exclude a possibility that there is some critical value for such a high $P_{sc}$ which could have caused the drop. 

\begin{figure}
\begin{center}
\includegraphics[width=0.3\textwidth,angle=270]{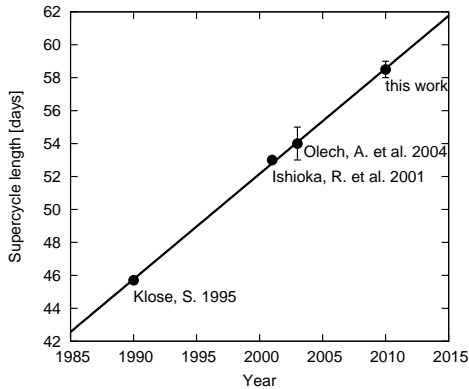}
\caption{
Plot from \citet{2012MOH} which shows the increasing supercycle length of IX~Dra during the past twenty years. 
The data points are taken from the literature. 
Uncertainties are given when available.
The line represents the best fit to the data.
The corresponding rate of increase of the period is $\dot{P} = 1.8 \times 10^{-3}$.}
\label{fig:Tsc}
\end{center}
\end{figure}

For completeness, in Fig.~\ref{fig:Tsc} we additionally present the plot from our previous publication \citep{2012MOH}, which shows the increasing supercycle length of another active ER~UMa-type star, IX Draconis.
This plot indicates that the rate of the period change is $\dot{P} = 1.8 \times 10^{-3}$, see Fig.\ref{fig:Tsc}. 
This result is based on four independent measurements from various papers from the literature \citep{2001Ishioka, 1995Klose, 2004Olech, 2012MOH}, which were obtained from short observational campaigns in each case. 
This star was in our original list of objects for the present study, however, the examination of its whole light curve was impossible since it was too poor to perform the ANOVA analysis. 

For all of our targets we found positive values of their period derivatives of supercycle lengths, $\dot{P}_{sc}$. They are presented in Table~\ref{tab-results}. 

\begin{table}
\caption{Rates of changes of $P_{sc}$ which correspond to the best fit to our measurements, presented in Fig.~\ref{fig-res-anova}, \ref{fig-res-constant}, and \ref{fig:Tsc}, together with respective ranges of $P_{sc}$ and \textit{HJD}.}
\begin{tabular}{c|c|c|c|c}
\hline
Object		& 	$\dot{P}_{sc}$					& $P_{sc}^{min}$ &  $P_{sc}^{max}$ & \textit{HJD} range \\
			&									& [days]		&	[days]		& [years] \\
\hline
DI~UMa 		&	$(4.3\pm9.6) \times 10^{-4} $ 			&	23.8	&44.7	&  30\\
ER~UMa 		&	$(12.7\pm 1.9) \times 10^{-4}$ 			&	42.7	&51.5	&  20\\
IX~Dra		&	$(17.5\pm 0.3) \times 10^{-4}$ 			&	45.7	&59.0	&  20\\
RZ~LMi 		&	$(5.0\pm 1.9) \times 10^{-4}$ 			&	18.3	&22.9	&  18\\
SS~UMi 		&	$(34.5\pm 5.8) \times 10^{-4} $ 		&	56.5	&108.3	&  29\\
V1159~Ori 	&	$(11.0\pm 4.9) \times 10^{-4} $ 		&	42.6	&55.1	&  19\\
V1504~Cyg	&	$(0.3\pm 5.4) \times 10^{-4}  $ 		&	110.6	&120.6	&  20\\
V344~Lyr	&	$(0.2\pm 2.5) \times 10^{-4}  $ 		&	112.9	&123.9	&  30\\
V503~Cyg 	&	$(0.6\pm 1.7) \times 10^{-3} $ 			&	112.1	&128.1	&  20\\
YZ~Cnc 		&	$(30.2\pm 1.4) \times 10^{-4} $ 		&	90.5	&149.5	&  51\\
\hline
\end{tabular}
\label{tab-results}
\end{table}

Only in two cases, i.e. V344~Lyr and V1504~Cyg, the $\dot{P}_{sc}$ values are so small that they seems to be constant. Within the uncertainties they could also be either positive or negative, thus the conclusive distinction is impossible here.
For two other objects, DI~UMa and V503~Cyg, it is also possible that within the uncertainties the values of $\dot{P}_{sc}$ are positive, constatnt, or negative. 
This is caused by the fact that measurements of their $P_{sc}$ show the highest dispersion. 
Nonetheless, for these two stars growing trends of their supercycle lengths are clearly visible in the Fig.~\ref{fig-res-anova}.

A comparison of these results with other examples known from the literature is satisfactory. 
They are in agreement at least with the order of magnitude. 
In the case of the ER~UMa star, we found two previously published values of its supercycle length derivative: 
$\dot{P}_{sc} \simeq 4 \times 10^{-3}$ \citep{1995Robertson}, 
and $\dot{P}_{sc} = 6.7(6) \times 10^{-4}$ \citep{2012Zemko}.
The first one is based on observations which were gathered between 1993-1995 ($\rm{HJD}\approx2449000$), so it is incomparable with our result.  
However, authors of this publication show that the mean supercycle length equals $42.95\pm0.06$~d, which is in a perfect agreement with the $P_{sc}$ for $\rm{HJD}\approx2449000$ in our Fig.~\ref{fig-res-anova}. 
On the other hand, the second result presented by \citet{2012Zemko} is based on analysis of a light curve with the time coverage of twenty years. Here the result is consistent with ours to the order of magnitude, and the small difference probably comes from the differences in the data sources and methods of analysis.
Another object, RZ~LMi, has also a published value of $\dot{P}_{sc} = -1.7 \times 10^{-3}$ \citep{1995Robertson}. This one is based on observations taken between 1992-1995 and is completely discrepant with our result. This is probably caused by the fact that this observing time was not long enough to reach any other conclusion, and it perhaps reflects some minor fluctuations of the supercycle length.
\citet{2000Kato} found the mean supercycle for SS~UMi to be equal to $84.7$~d for the observations time $\rm{HJD}\approx2451000$, which is again in a perfect agreement with our Fig.~\ref{fig-res-anova}. 
The same is for V1159~Ori, with $P_{sc}=45$~d around $\rm{HJD}\approx2450000$, as presented by \citet{1999Szkody}.
In turn \citet{2005Antonyuk} showed for the five years spread of observations that the supercycle for V1504~Cyg seems to stay more or less constant, apart from short time scale fluctuations. 

Our examination of supercycle lengths' behaviour on long time scales demonstrates that all of the analysed active SU~UMas have a positive value of their superoutbursts period derivative, and eight out of ten objects have a very clearly increasing trend in their $P_{sc}$ evolution. This behaviour seems to be general for this type of stars. 
The fact that for most of these sources there is a clear evidence of increasing supercycle lengths means that their mean mass transfer rates has been constantly decreasing during the past decades. 
This scenario of evolution is in agreement with results for extremely interesting ER~UMa-type object, BK~Lyncis, presented by \citet{2012Patterson}. In this publication the authors showed a hypothesis of the evolutionary path of this star as its ER~UMa stage is a transient phase of evolution, preceded by classical nova and novalike phases. If we assumed that this is true for all active SU~UMas of our sample, this would mean that these objects are constantly fading since their classical nova eruptions. Considering the fact that they are rather diverse in terms of their orbital periods, i.e. evolutionary stages (see Fig.~\ref{fig-evol}), even though they are all below the period gap, we can try to predict time scales of their further steps of evolution. 
All of our objects are extremely active as for the SU~UMa class. 
The most active of them are sometimes distinguished as a separate class of ER~UMa-type, with $P_{sc}\approx50~\rm{d}$. They are believed to be typical SU~UMas with only higher mass transfer rates (DI~UMa, ER~UMa, IX~Dra, RZ~LMi, and V1159~Ori from our sample). 
The rest of the sample are representatives of the SU~UMa class of stars but still with extremely short supercycles (SS~UMi, V1504~Cyg, V344~Lyr, V503~Cyg, and YZ~Cnc). 
A typical supercycle length for a star of SU~UMa type is equal to a few hundred days and for the WZ~Sge class it is of the order of decades \citep{2001Hellier}. 
We adopted these standard values to find the expected time scales in which our objects will reach next levels of evolution. 
For all of the stars we estimate the time which is needed to become SU~UMa-type object with $P_{sc}=300~\rm{d}$. 
For the two period bouncers in our sample, DI~UMa and IX~Dra, we also give times needed to reach the next step that is becoming a WZ~Sge star with $P_{sc}=10~\rm{yr}$. 
Results of this estimate are shown in Table~\ref{tab-nextSteps}.

\begin{table}
\caption{Estimates of time scales of next steps of evolution of our objects under the assumption of the constant growth of their supercycle lengths. The second column shows in how many years a star will become a typical SU~UMa star with $P_{sc}=300~\rm{d}$. The third column shows in how many years period bouncers will attain the stage of WZ~Sge star with $P_{sc}=10~\rm{yr}$.}
\begin{tabular}{c|c|c}
\hline
	Object	&	SU UMa stage 		 &	WZ Sge stage 	\\
			& ($P_{sc}=300~\rm{d}$)	& ($P_{sc}=10~\rm{yr}$) \\
\hline
DI~UMa 		&		1700	&	23000	\\
ER~UMa 		&		500	&	--			\\
IX~Dra 		&		400	&	6000 		\\
RZ~LMi 		&		1500	&	--		\\
SS~UMi 		&		200	&	--			\\
V1159~Ori 	&		600	&	--			\\
V1504 Cyg	&		20000	&	--		\\
V344 Lyr	&		33000	&	--		\\
V503~Cyg 	&		800	&	--			\\
YZ~Cnc 		&		100	&	--			\\
\hline
\end{tabular}	
\label{tab-nextSteps}
\end{table}

\section{Conclusions}
\label{sec-conc}
We analysed a set of photometric data covering last decades of observations for the most active SU~UMa stars in order to study changes of their supercycle lengths.
Despite the common feature of the selected objects, which is the highest activity in this class of stars, 
they are diverse in terms of their orbital periods below the period gap. 
For all of the analysed objects we found positive values of period derivatives of their supercycle lengths. 
There are some subtle fluctuations for short time scales in the behaviour of $P_{sc}$, but the general trend is the same in each case. 
Increasing supercycle lengths mean that the mean mass transfer rates have been decreasing for these objects over the last decades. This is in agreement with the scenario of the evolution of BK~Lyn presented by \citet{2012Patterson}, which seems to be a general case. This phenomenon is important in the context of evolution of such systems.

\section*{Acknowledgments}
\small{
We acknowledge with thanks the variable star observations 
from the AAVSO International Database, operated in USA,
the AFOEV database, operated at CDS, France,
the BAAVSS database, operated in UK,
and the Variable Star and Exoplanet Section of the Czech Astronomical Society, 
contributed by observers worldwide and used in this research.
This paper includes data collected by the Kepler mission. Funding for the Kepler mission is provided by the NASA Science Mission directorate.
Some of the data presented in this paper were obtained from the Mikulski Archive for Space Telescopes (MAST). STScI is operated by the Association of Universities for Research in Astronomy, Inc., under NASA contract NAS5-26555. Support for MAST for non-HST data is provided by the NASA Office of Space Science via grant NNX09AF08G and by other grants and contracts.
This research has made use of the Simbad database, operated at CDS, Strasbourg, France.
The project was supported by Polish National Science Center grant awarded by decision number DEC-2011/03/N/ST9/03289.

\bibliographystyle{mn2e} 
\bibliography{Psc_bib}
}

\bsp

\end{document}